\documentclass[aps,prl,twocolumn,
10pt,
superscriptaddress,preprintnumbers,showpacs,nofootinbib,
a4paper
]{revtex4-2}

\usepackage{graphicx}
\usepackage{siunitx}
\usepackage{upgreek}
\usepackage{tikz}
\usetikzlibrary{math}
\usepackage[linkcolor=blue,citecolor=blue,urlcolor=blue,colorlinks=true]{hyperref}
\usepackage[columnwise]{lineno}

\begin{document}


\title{Measurement of the Electron capture of \textsuperscript{76}As into the first excited state of \textsuperscript{76}Ge}

\author{Hans~F.~R.~Hoffmann}
\email[]{hans\_fritz\_rudolf.hoffmann@tu-dresden.de}
\author{Bj\"{o}rn Lehnert}
\author{Kai Zuber}
\affiliation{Institut f\"ur Kern- und Teilchenphysik (IKTP), Technische Universit\"at Dresden, Dresden, Germany}

\date{\today}

\begin{abstract}
The neutrinoless double beta decay of $^{76}$Ge is searched for in the large-scale experiment LEGEND. The measurement of the half-life of this process would give access to the neutrino mass using the nuclear matrix element. 
Experimentally the contribution of the $^{76}$As ground state to the nuclear matrix element can be investigated via the branching ratios of its $\beta^-$ and electron capture decay. 
While energetically, the electron capture of $^{76}$As into the first excited state of $^{76}$Ge is possible and was measured once before this work, 
the electron capture into the $^{76}$Ge ground state was not observed yet. 

The present study investigates the branching of $^{76}$As that is produced via $^{75}$As(n,$\gamma$) on a thin As$_2$O$_3$ sample. A silicon drift detector measures characteristic X-rays emitted by the germanium atoms caused by an inner vacancy after the electron capture. A high-purity germanium detector 
is used to measure the \SI{562.9}{\keV} $\gamma$-rays emitted after electron capture into the excited state. Investigation of coincident signals in both detectors leads to the branching ratio of the $^{76}$As electron capture into the first excited state of $^{76}$Ge of $\nu_{\mathrm{EC}^\ast} = (0.0572 \pm 0.0029 (\mathrm{stat.}) \pm 0.0074(\mathrm{syst.}))\%$. This is the first measurement with the full uncertainty budget quantified. 

\end{abstract}


\maketitle


\section{Introduction}
The neutrinoless double beta decay ($0\nu\beta\beta$) is a beyond standard model process violating lepton number conservation by 2 units that has not been measured so far \cite{Dolinski2019},\cite{Agostini2023}. 
Several large scale experiments are searching for it in various candidate isotopes \cite{Agostini2023}. 
The \textsc{Legend} experiment \cite{LEGEND} is pursuing the promising approach searching for the $0\nu\beta\beta$ decay in $^{76}$Ge. In a source equals detector approach, high-purity germanium (HPGe) detectors enriched in $^{76}$Ge are operated quasi background free \cite{Burlac2025}. 

A measurement of the neutrinoless double beta decay would prove the Majorana nature of the neutrino opening channels for the explanation of the matter-antimatter asymmetry in the universe. Furthermore, the measured half-life gives access to the (Majorana) neutrino mass on an absolute scale assuming light Majorana neutrino exchange \cite{SuhonenCivitarese2008} as given in equation \ref{eq:m_nu}. 
\begin{equation}
    \left(T_{1/2}^{0\nu}\right)^{-1} = G_{0\nu}\left(Q_{\beta\beta},Z\right) \cdot \left|M_{0\nu}\right|^2 \cdot \left(\frac{\left<m_\nu\right>}{m_e}\right)^2
    \label{eq:m_nu}
\end{equation}
Various theoretical models for the nuclear matrix element (NME) $\left|M_{0\nu}\right|^2$ exist \cite{Grabmayr2025}. 
Measurements of decay branches of the intermediate nucleus provide useful benchmarks for nuclear-structure calculations that are also used in NME calculations.

\begin{figure}
    \includegraphics[width=\linewidth]{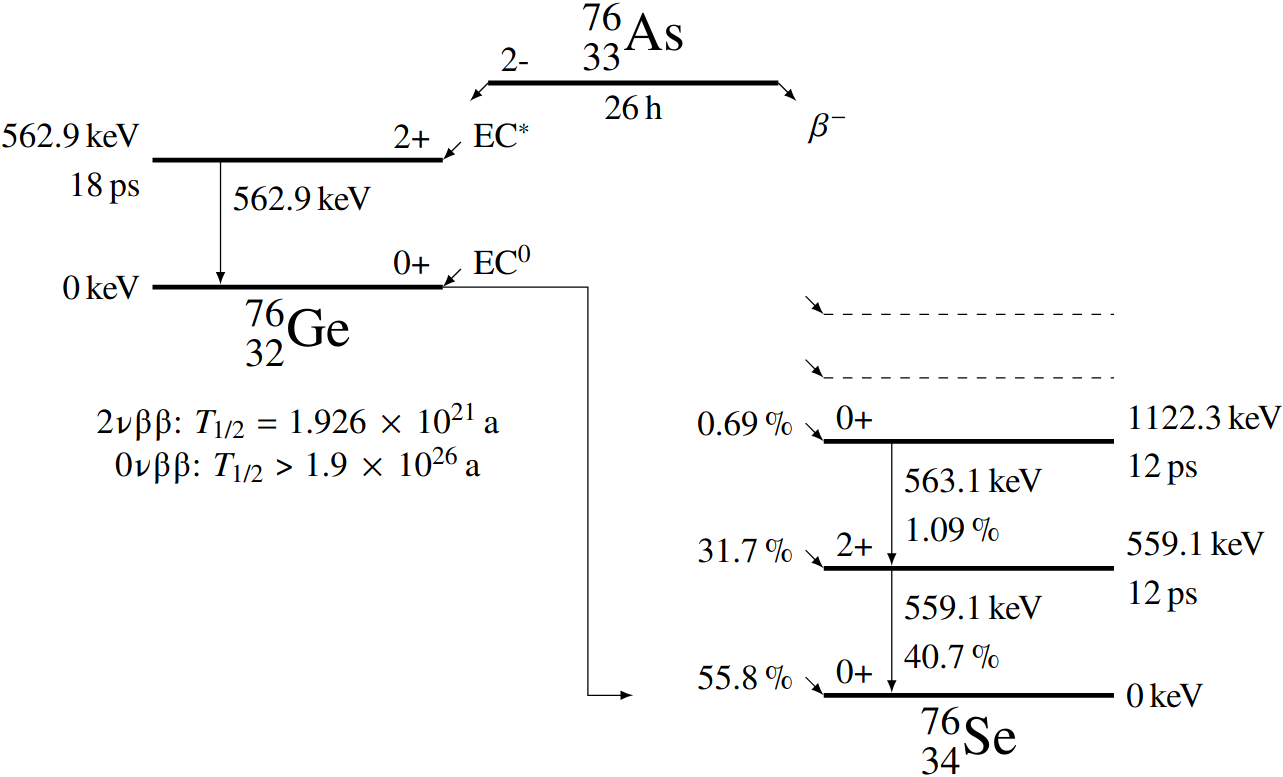}
    \caption{Decay scheme of \textsuperscript{76}As with selected $\gamma$-ray lines with energy close to the EC$^{\ast}$ $\gamma$-ray energy $562.9\,$keV.}
    \label{fig:decay-scheme}
\end{figure}

The intermediate nucleus in the $0\nu\beta\beta$ of $^{76}$Ge is $^{76}$As, a radioactive isotope with a half-life of $(26.254 \pm 0.011)\,$h. 
A decay scheme is shown in figure~\ref{fig:decay-scheme}. 
It decays by close to \SI{100}{\percent} via $\beta^{-}$ into states of $^{76}$Se, 
$(55.8\pm0.3)\%$ 
of which directly into the ground state~\cite{Singh2024}. 
Both $\beta^{-}$ and electron capture (EC) $Q$-values are positive, $(2960.6 \pm 0.9)\,$keV and $(921.5 \pm 0.9)\,$keV, respectively~\cite{Meng2020}. 
Energetically, the EC into the ground state (0$^+$) and first excited state (2$^+$) of $^{76}$Ge is possible. 
The 2$^-$ to 2$^+$ EC into the excited state (EC$^\ast$) is 
a 1st-forbidden non-unique EC transition which is interesting as such. 
It was measured once before, resulting in a branching ratio of $\approx$0.027\% \cite{Domula2014} 
corresponding to a $\mathrm{log}\,ft$ value of 7.34 \cite{Turkat2023}. 
The $\mathrm{log}\,ft$ value is a measure of transition strength depending on the partial half life, phase space factors and nuclear structure effects. 
Higher branching ratios are associated with lower $\mathrm{log}\,ft$ values.
The EC into the 0$^+$ ground state (EC$^0$) has not been measured yet. 
Theoretical calculations predict $\mathrm{log}\,ft$ = 8.78 for EC$^0$ and $\mathrm{log}\,ft$ = 7.05 for EC$^\ast$ leading to branching ratios $\approx\!0.03\%$ for both channels \cite{SuhonenPriv}. 

The experimental signature of the EC$^\ast$ is a $562.917(23)\,$keV 
$\gamma$-ray in coincidence with a characteristic Ge X-ray. The physical time difference of $\gamma$-ray and X-ray emission is on the fs time scale.

\section{Measurement}
In order to measure the branching ratio of the EC$^\ast$, a  sample is activated. In a coincidence counting setup, the simultaneously emitted 562.9\,keV $\gamma$ ray and the $K_\alpha$ X-ray are measured. 

The sample is made of $(195.2\!\pm\!1.0)\,$mg As$_2$O$_3$ powder arranged as a disk with a diameter of 22\,mm resulting in an average areal density of 
$(0.05135\!\pm\!0.00026)\,\mathrm{g}\,\mathrm{cm}^{-2}$. 
The powder is held in a sample container with an outer diameter of \SI{24}{mm} and a total height of \SI{8}{\mm}. The sample container covers the 
powder layer with \SI{1}{\mm} of polycarbonate. 

The isotope $^{76}$As is produced in the sample via $^{75}\mathrm{As(n,} \gamma\mathrm{)}$ using the thermal neutrons 
of the AKR-2 research reactor 
\cite{Rodriguez2026} at Technische Universit\"{a}t Dresden. The sample is activated in the central activation channel of the reactor inside the fission zone. The reactor 
provides a neutron flux density of $2.5\!\times\!10^7\,\mathrm{cm}^{-2}\mathrm{s}^{-1}$. 
The sample is activated for \SI{8}{\hour} in the reactor and then transported to the counting setup of the experiment within \SI{1}{\hour} and measured for a few days. This cycle was repeated eleven times. 

The counting setup is located in the shallow-underground laboratory Felsenkeller Dresden \cite{Bemmerer2025} and
consists of the HPGe called TU2 
and the silicon drift detector (SDD) TU3. 
\begin{figure}
    \centering
    \begin{tikzpicture}[scale = 0.1]
        \newlength{\w}
        \setlength{\w}{0.75pt} 
        \newlength{\dGe}
        \setlength{\dGe}{10mm} 
        \newlength{\rBh}
        \setlength{\rBh}{10.5mm} 
        \newlength{\lBh}
        \setlength{\lBh}{40mm} 
        \newlength{\rGe}
        \setlength{\rGe}{36mm} 
        \newlength{\lGe}
        \setlength{\lGe}{65mm} 
        \newlength{\rCryo}
        \setlength{\rCryo}{44.5mm} 
        \newlength{\lCryo}
        \setlength{\lCryo}{160mm} 
        \newlength{\rinCryo}
        \setlength{\rinCryo}{8mm} 
        \newlength{\rSc}
        \setlength{\rSc}{12mm} 
        \newlength{\rinSc}
        \setlength{\rinSc}{9mm} 
        \newlength{\zSc}
        \setlength{\zSc}{8mm} 
        \newlength{\dSc}
        \setlength{\dSc}{1mm} 
        \newlength{\rAs}
        \setlength{\rAs}{11mm} 
        \newlength{\dAs}
        \setlength{\dAs}{0.5mm} 
        \newlength{\zD}
        \setlength{\zD}{440mm} 
        \newlength{\nX}
        \setlength{\nX}{108mm} 
        \newlength{\rXn}
        \setlength{\rXn}{12.5mm} 
        \newlength{\tX}
        \setlength{\tX}{2mm} 
        \newlength{\rX}
        \setlength{\rX}{6mm} 
        \newlength{\dX}
        \setlength{\dX}{0.45mm} 
        \newlength{\dColl}
        \setlength{\dColl}{0.45mm} 
        \newlength{\rColl}
        \setlength{\rColl}{5mm} 
        \newlength{\RColl}
        \setlength{\RColl}{7.5mm} 
        \newlength{\rXh}
        \setlength{\rXh}{10.5mm} 
        \newlength{\zXh}
        \setlength{\zXh}{9.2mm} 
        \newlength{\rbX}
        \setlength{\rbX}{39mm} 
        \newlength{\bX}
        \setlength{\bX}{145mm} 
        \newlength{\zX}
        \setlength{\zX}{13mm} 
        \newlength{\dCu}
        \setlength{\dCu}{50mm} 
        \newlength{\rCu}
        \setlength{\rCu}{100mm} 
        \newlength{\ztCu}
        \setlength{\ztCu}{600mm} 
        \newlength{\zbCu}
        \setlength{\zbCu}{150mm} 
        \newlength{\rPb}
        \setlength{\rPb}{250mm} 
        \newlength{\zPb}
        \setlength{\zPb}{800mm} 
        \newlength{\dRn}
        \setlength{\dRn}{10mm} 
        \newlength{\rRn}
        \setlength{\rRn}{125mm} 
        \newlength{\dHp}
        \setlength{\dHp}{10mm} 
        \newlength{\rHp}
        \setlength{\rHp}{170mm} 
        \newlength{\dPbCap}
        \setlength{\dPbCap}{50mm} 
        \newlength{\rPbCap}
        \setlength{\rPbCap}{150mm} 
        \draw [fill=red!25!white, line width = \w] (\rCryo, -\lCryo) -- (\rCryo, 0) -- (\rinCryo, 0) -- (\rinCryo, -\lBh) -- (-\rinCryo, -\lBh) -- (-\rinCryo, 0) -- (-\rCryo, 0) -- (-\rCryo, -\lCryo) -- cycle;
        \filldraw [draw=black, fill=red, line width=0.1pt] (\rGe, -\lGe-\dGe) -- (\rGe, -\dGe) -- (\rBh, -\dGe) -- (\rBh, -\lBh-\dGe) -- (-\rBh, -\lBh-\dGe) -- (-\rBh, -\dGe) -- (-\rGe, -\dGe) -- (-\rGe, -\lGe-\dGe) -- cycle;
        \draw [draw=black, fill=teal!20!white, line width=\w] (\rSc, 0) -- (\rSc, \zSc) -- (-\rSc, \zSc) -- (-\rSc, 0) -- (-\rinSc, 0) -- (-\rinSc, \zSc-2.0\dSc-\dAs) -- (\rinSc, \zSc-2.0\dSc-\dAs) -- (\rinSc, 0) -- cycle; 
        \draw [draw=black, fill=teal, line width=0.1pt] (-\rAs, \zSc-\dSc-\dAs) rectangle (\rAs, \zSc-\dSc); 
        \begin{scope} [yshift=\zX]
            \draw [line width=\w, fill=blue!20!white] (-\rXh,0) rectangle (\rXh,\zXh); 
            \draw [line width=\w, fill=blue!20!white] (-\rXn,\zXh) rectangle (\rXn,\nX); 
            \draw [line width=\w, fill=blue!20!white] (-\rbX,\nX) rectangle (\rbX,\nX+\bX); 
            \draw [fill=blue, draw=black, line width=0.1pt] (-\rX, \tX) rectangle (\rX, \tX+\dX); 
        \end{scope}
        \begin{scope}[yshift=-\zD]
            \filldraw [draw=black, fill=gray, line width=\w] (\rCu, \zbCu) -- (\rCu, \ztCu) -- (\rCu-\dCu, \ztCu) -- (\rCu-\dCu, \zbCu+\dCu) -- (-\rCu+\dCu, \zbCu+\dCu) -- (-\rCu+\dCu, \ztCu) -- (-\rCu, \ztCu) -- (-\rCu, \zbCu) -- cycle;
            \filldraw [draw=black, fill=lightgray, line width = \w] (\rPb, 0) -- (\rPb, \zPb) -- (\rCu, \zPb) -- (\rCu, \zbCu) -- (-\rCu, \zbCu) -- (-\rCu, \zPb) -- (-\rPb, \zPb) -- (-\rPb, 0) -- cycle;
            \draw [fill=gray, line width = \w] (\rPb+\dRn, -\dRn) -- (\rPb+\dRn, \zPb+\dRn) -- (\rRn, \zPb+\dRn) -- (\rRn, \zPb) -- (\rPb, \zPb) -- (\rPb, 0) -- (-\rPb, 0) -- (-\rPb, \zPb) -- (-\rRn, \zPb) -- (-\rRn, \zPb+\dRn) -- (-\rPb-\dRn, \zPb+\dRn) -- (-\rPb-\dRn, -\dRn) -- cycle;
        \end{scope}
        \begin{scope}[yshift=-\zD+\zPb+\dRn+\dHp]
            \draw [fill=gray, line width=\w] (-\rHp, -\dHp) rectangle (\rHp, 0);
            \filldraw [draw=black, fill=lightgray, line width = \w] (-\rPbCap, 0) rectangle (\rPbCap, \dPbCap);
        \end{scope}
        \newlength{\rC}
        \setlength{\rC}{20mm} 
        \newlength{\oC}
        \setlength{\oC}{300mm} 
        \tikzmath{
            \fC = 8;
            \aC = \oC+\fC*\rC;
            \snb = (\fC-1)*\rC/\aC;
            \csb = 0.95256097; 
        }
        \draw [line width=0.1pt, draw=darkgray] (0,0) circle  (\rC);
        \draw [line width=0.1pt, draw=darkgray] (-\snb*\rC,\csb*\rC) -- (\aC-\snb*\rC*\fC,\csb*\rC*\fC);
        \draw [line width=0.1pt, draw=darkgray] (-\snb*\rC,-\csb*\rC) -- (\aC-\snb*\rC*\fC,-\csb*\rC*\fC);
        \begin{scope}[xshift=\aC, scale = \fC]
            \draw [line width=0.1pt, draw=darkgray] (0,0) circle  (\rC);
            \clip (0,0) circle[radius=\rC];
        \draw [fill=red!25!white, line width = \w] (\rCryo, -\lCryo) -- (\rCryo, 0) -- (\rinCryo, 0) -- (\rinCryo, -\lBh) -- (-\rinCryo, -\lBh) -- (-\rinCryo, 0) -- (-\rCryo, 0) -- (-\rCryo, -\lCryo) -- cycle;
        \filldraw [draw=black, fill=red, line width=0.1pt] (\rGe, -\lGe-\dGe) -- (\rGe, -\dGe) -- (\rBh, -\dGe) -- (\rBh, -\lBh-\dGe) -- (-\rBh, -\lBh-\dGe) -- (-\rBh, -\dGe) -- (-\rGe, -\dGe) -- (-\rGe, -\lGe-\dGe) -- cycle;
        \draw [draw=black, fill=teal!20!white, line width=\w] (\rSc, 0) -- (\rSc, \zSc) -- (-\rSc, \zSc) -- (-\rSc, 0) -- (-\rinSc, 0) -- (-\rinSc, \zSc-2.0\dSc-\dAs) -- (\rinSc, \zSc-2.0\dSc-\dAs) -- (\rinSc, 0) -- cycle; 
        \draw [draw=black, fill=teal, line width=0.1pt] (-\rAs, \zSc-\dSc-\dAs) rectangle (\rAs, \zSc-\dSc); 
        \begin{scope} [yshift=\zX]
            \draw [line width=\w, fill=blue!20!white] (-\rXh,0) rectangle (\rXh,\zXh); 
            \draw [fill=blue, draw=black, line width=0.1pt] (-\rX, \tX) rectangle (\rX, \tX+\dX); 
        \end{scope}
        \end{scope}
        \newlength{\rLbl}
        \setlength{\rLbl}{300mm} 
        \node [anchor=center, blue] at (460mm, 280mm) {X-ray SDD};
        \node [anchor=center, teal] at (460mm, 240mm) {As sample};
        \node [anchor=center, red] at (460mm, 200mm) {HPGe};
        \draw [line width = 0.1pt] (75mm, -265mm) -- (\rLbl, -\zD+\zbCu+0.5\dCu) node [anchor=west] {Cu shielding};
        \draw [line width = 0.1pt] (175mm, -315mm) -- (\rLbl, -315mm) node [anchor=west] {Pb shielding};
        \draw [line width = 0.1pt] (255mm, -365mm) -- (\rLbl, -365mm) node [anchor=west] {Anti-Rn box};
    \end{tikzpicture}
    \caption{Schematic illustration of the experimental setup. The activated 
    As$_2$O$_3$ powder is arranged as a thin layer (137\,$\upmu$m) inside the sample container covering it with \SI{1}{\mm} of polycarbonate. 
    The sample is placed on the cryostat end cap of the HPGe. The HPGe measures $\gamma$-rays emitted in the decay of $^{76}$As. A silicon drift detector measures the characteristic Ge X-rays emitted in the electron capture of $^{76}$As. The setup is covered from natural radiation by a Cu and Pb shielding and an anti-Rn box.}
    \label{fig:setup-scheme}
\end{figure}
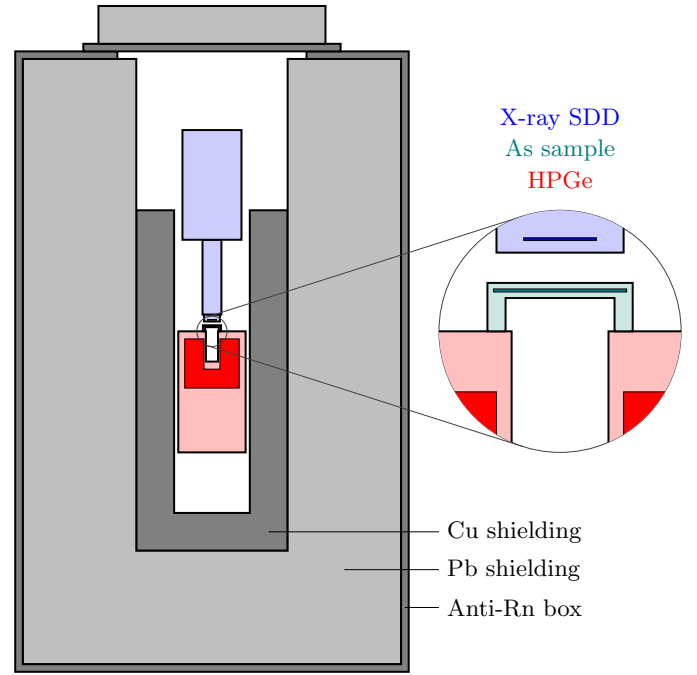
The HPGe is a Canberra GSW200 small anode germanium well detector with a thin lithium diffused contact inside the well. 
The germanium crystal is \SI{72}{\mm} in diameter and \SI{65}{\mm} in length with a well on the front side of \SI{40}{\mm} depth and \SI{21}{\mm} in diameter. 
The active volume is \SI{218}{\cm^3} and the relative efficiency is \SI{53.5}{\percent}. 
The HPGe has an excellent energy resolution of $\mathrm{FWHM} = 0.79\,$keV at $E_\gamma = 562.9\,$keV. 
The X-ray 
sensor is a Vitus H80 cube, a $d = \SI{450}{\micro\m}$, $A = \SI{109}{\mm\squared}$ silicon chip with a multi-layer on-chip collimator defining the open area of $\SI{80}{\mm\squared}$. The entrance window is \SI{25}{\um} of Beryllium. 
The Si sensor is brought to the operating temperature of \SI{-35}{\degreeCelsius} by Peltier cooling. 
The X-ray detector implements a reset preamplifier. The output signal rises with 60\,V/$\upmu$s and is reset to 0 when 1.5\,V is reached. Signals appear as steps in the voltage output over time. 
The FWHM resolution achieved with the X-ray detector in this experiment is 0.213\,keV at 10\,keV. 
A scheme of the counting setup arrangement is shown in figure~\ref{fig:setup-scheme}. 
The activated As sample is placed centrally on the cryostat end cap of the HPGe. 
The SDD is aligned facing downwards with the entrance window positioned \SI{5}{\mm} above the sample. 
The setup is inside the \SI{5}{\cm} Cu and \SI{15}{\cm} Pb shielding of the HPGe with the top parts of the shielding removed, making space for the X-ray detector, and covered again by \SI{5}{\cm} of Pb. 

Data acquisition (DAQ) is done using a CAEN DT5725S multi-channel digitizer. Data of both detectors are stored in list mode event-wise with their time stamp, reconstructed pulse height and the input pulse shape. 
An incorrect digitizer dynamic-range configuration resulted in partial truncation of the X-ray detector preamplifier reset signal. 
This introduces a significant dead time. Signal X-rays can only be measured when both the preamplifier signal before and after the input pulse are inside the DAQ's dynamic range. 

With measurement durations of the activated sample up to five $^{76}$As half-life periods, a wide spread of count rates occurred. Physical input count rates ranged from 113 to 5400\,s$^{-1}$ for the HPGe and from 2 to 76\,s$^{-1}$ for the X-ray detector. 
\begin{figure*}
    \includegraphics[width=\textwidth]{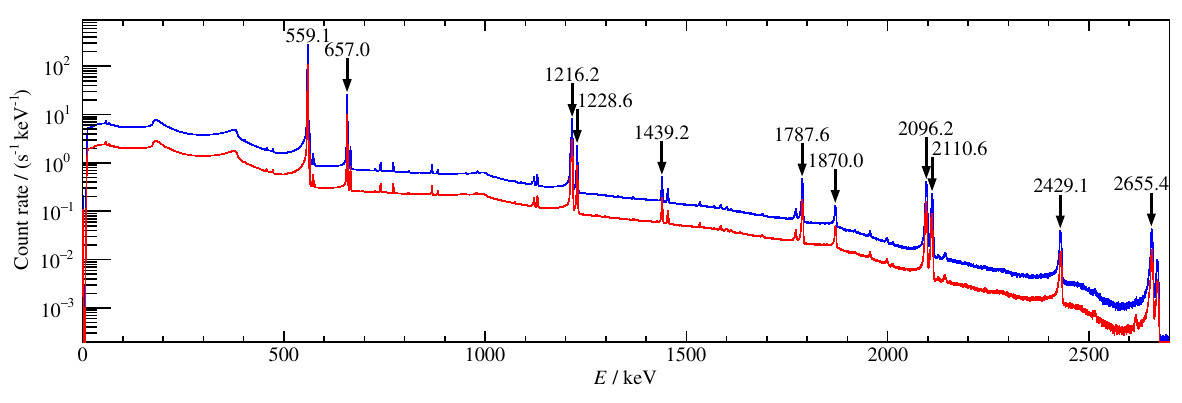}
    \caption{HPGe $\gamma$-ray energy spectrum for all measurement runs before coincidence cuts. The experimental data is divided into two data sets, corresponding to higher (blue) and lower (red) count rate. The threshold count rate is set such that both sets have the same \SI{559.1}{\keV} reference counts. 
    }
    \label{fig:TU2_spectrum_halves}
\end{figure*}
\section{Analysis}
The aim is to measure the branching ratio in a relative approach. The number of EC$^\ast$ decays is found by coincident detection of the 562.9\,keV $\gamma$-ray in the HPGe and the 9.9\,keV Ge X-ray in the SDD. 
A difficulty arises from the 
563.1\,keV $\gamma$-ray line from the $\beta^-$ decay of $^{76}$As overlapping with the signal line. 
The number of $^{76}$As decays is estimated using the detection of the \SI{559.1}{\keV} $\gamma$-ray. 

For the HPGe energy calibration, eight prominent $\gamma$-ray lines 
from the $\beta^-$ decay of $^{76}$As are used. 
The calibrated HPGe spectrum is shown in figure~\ref{fig:TU2_spectrum_halves}. 
For the SDD energy calibration, the As $K_\alpha$ and $K_\beta$ lines are used. 
\begin{figure}
    \includegraphics[width=\linewidth]{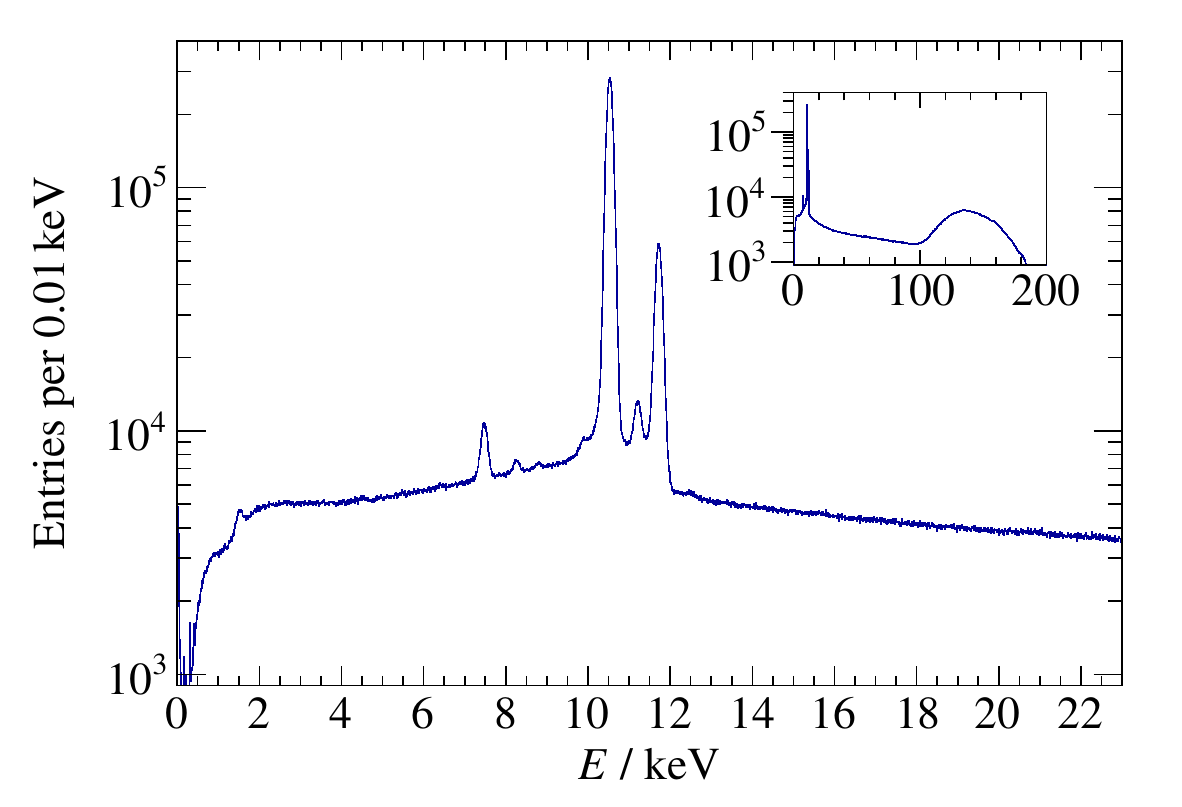}
    \caption{Calibrated X-ray energy spectrum before coincidence cuts summed over all runs. }
    \label{fig:TU3}
\end{figure}
The calibrated X-ray spectrum shown in figure~\ref{fig:TU3} is dominated by the X-ray peaks labeled in figure~\ref{fig:CoincProj} and a continuum, both from the decay of $^{76}$As. The enhancement in the continuum above 100\,keV is due to $\beta^-$ particles. The peak at 7.5\,keV is the Ni $K_\alpha$ line due to X-ray fluorescence in the on-chip collimator. 

\subsection{Coincidence events}
The signal cannot be obtained by a simple cut around the expected coincidence energy because the region of interest contains both accidental coincidences and true $\beta$-decay coincidences. A background estimate therefore has to account for both the $\gamma$-ray and X-ray energy distributions. A two-dimensional sideband method is applied for the quantification of the background. 

\begin{figure}[h]
    \includegraphics[width=\linewidth]{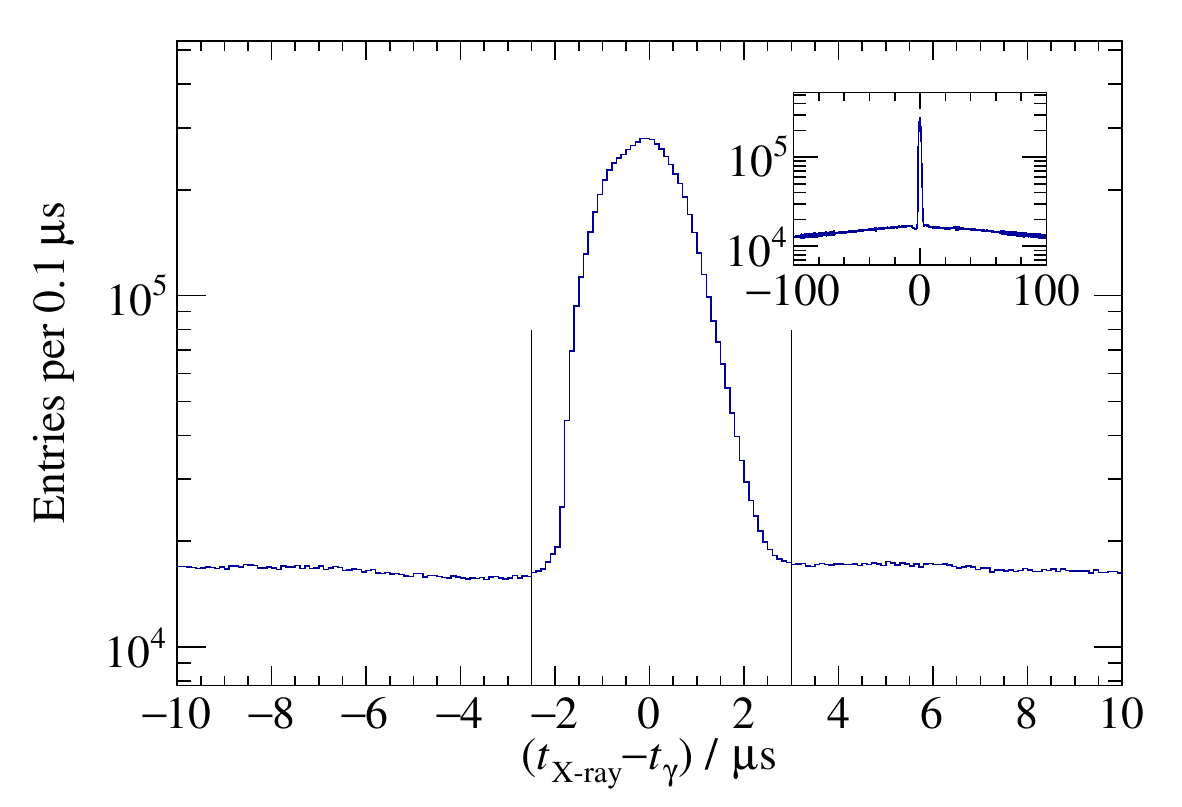}
    \caption{Time difference $t_\text{X-ray} - t_\gamma$ between subsequent events in both HPGe and X-ray detector for the total data set. The time difference limits $\SI{-2.5}{\us}$
     and $\SI{3}{\us}$ are shown as vertical lines. \SI{0.1}{\micro\second} binning. Inlet: Same data, shown on a larger time scale with the same binning. 
    }
    \label{fig:dt}
\end{figure}
The time difference distribution of consecutive events in different detectors 
in figure~\ref{fig:dt} 
shows a true coincidence peak upon a background. The inlet shows the same data for a wider time difference range. The background is exponentially distributed as expected for random coincidences. 
Random coincidence events are rejected by the time difference cut condition $\SI{-2.5}{\us} < t_{\text{X-ray}} - t_\gamma < \SI{3}{\us}$. 
The amount of random coincidences being contained in the coincidence time window depends on the sample activity which is varying during the measurement. 
Estimated from background bands at both sides of the peak, for the complete dataset, a fraction of 12\% of the events remaining after the coincidence cut are random coincidences. 

\begin{figure*}
    \centering
    \includegraphics[width=\linewidth]{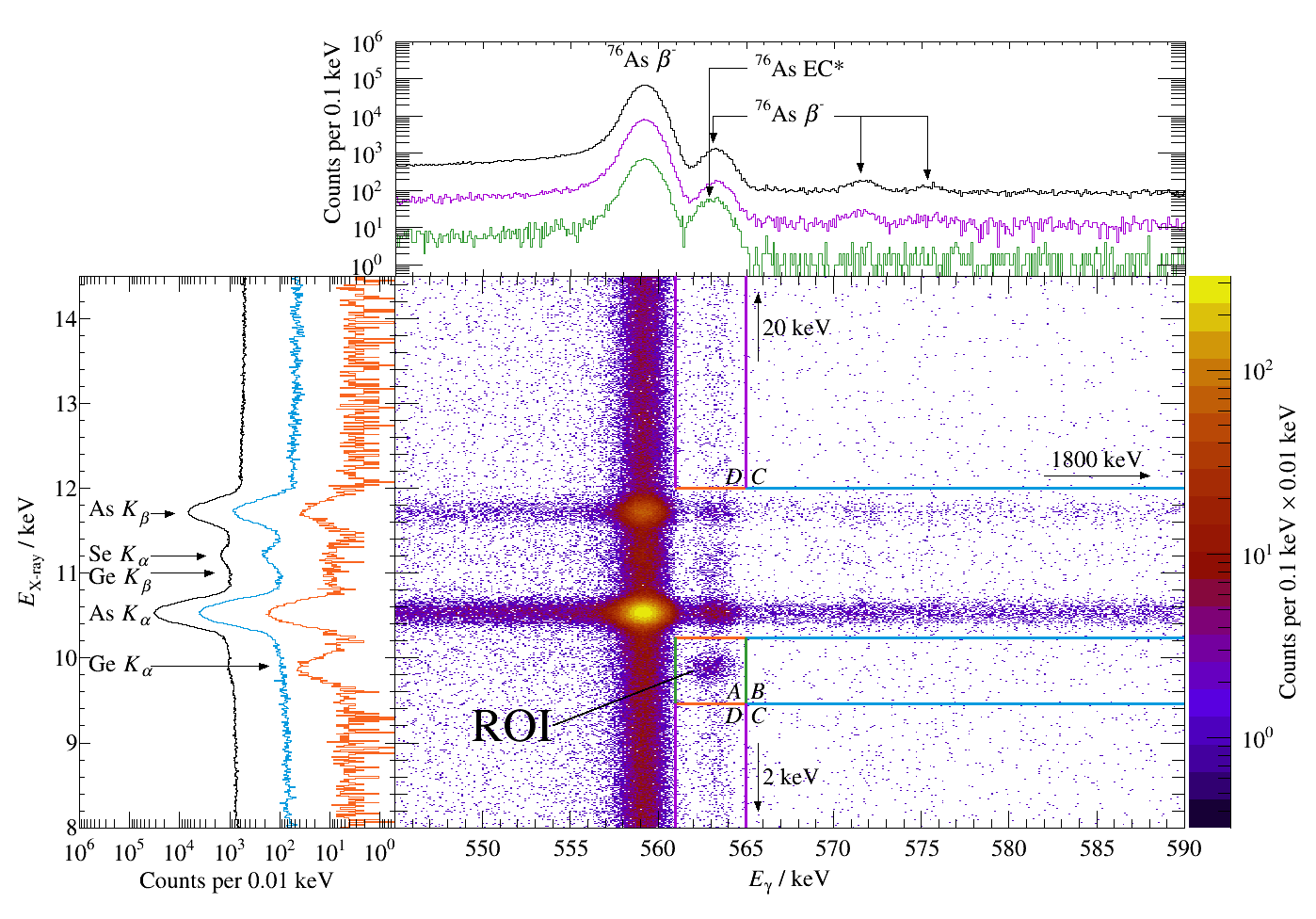}
    \caption{Reconstructed X-ray and $\gamma$-ray energies of coincident event pairs with the time difference condition $\SI{-2.5}{\us} < t_{\text{X-ray}} - t_\gamma < \SI{3}{\us}$. 
    Energy depositions in the HPGe are 559.1\,keV and 563.1\,keV full-energy events, their respective multi-scatter Compton continua and Compton continua of higher-energy $\gamma$-ray lines. In the X-ray detector are full-energy depositions of As $K_\alpha$ (10.5\,keV), Se $K_\alpha$ (11.2\,keV), As $K_\beta$ (11.7\,keV) and Se $K_\beta$ (12.5\,keV) X-rays and a continuous background. 
    The number of coincident events is evaluated in the region of interest (ROI) marked as $A$. The background is estimated in regions $B$, $C$ and $D$. The summation X-ray and $\gamma$-ray energy intervals extend beyond the shown section to 2\,keV, 20\,keV and 1800\,keV, respectively.
    The 
    $\gamma$-ray energy
    projections shown are for all coincident events in black and colored for the signal and background X-ray energy cuts, respectively, and analogously for the X-ray projections. }
    \label{fig:CoincProj}
\end{figure*}

Figure~\ref{fig:CoincProj} 
shows the reconstructed energies in both detectors for event pairs fulfilling the time difference condition. 

The EC$^\ast$ signals are expected at $E_\gamma = \SI{562.9}{\keV}$ and $E_{\text{X-ray}} = \SI{9.9}{\keV}$. Nearby features are dominated by true coincidences from the $^{76}$As $\beta^-$ branch. 
$\beta^-$ electrons induce fluorescence in the sample material leading to the detection of As $K_\alpha$ and $K_\beta$ X-rays in the SDD in coincidence with $\gamma$ rays being detected in the HPGe. Se $K_\alpha$ and $K_\beta$ X-rays are emitted by the $^{76}$As $\beta^-$ daughter atom, with the probabilities $(4.26\pm0.46)\!\times\!10^{-2}\,\%$ and $(7.0\pm1.0)\!\times\!10^{-3}\,\%$ respectively \cite{Schoenfeld1996}. 
Coincident detection of both a Se $K$ X-ray in the SDD and a $\gamma$-ray from the $\beta^-$ cascade makes another true coincident feature close to the region of interest. 
EC$^\ast$ decays can lead to coincident signals via $K_\beta$ emission and detection in the SDD and coincident detection of the 562.9\,keV $\gamma$-ray in the HPGe.  This leads to the peak at the Ge $K_\beta$ energy of 11.0\,keV in figure~\ref{fig:Stat_indep_cross_check}. With the Ge $K_\alpha$ signal peak being more distinct in the coincidence correlation plot and having better statistics, only the $K_\alpha$ X-rays are considered for the analysis. 
Random coincidences originate from two independent physical decays occurring sufficiently close in time to satisfy the coincidence condition. 
Consequently, the measured $\gamma$-ray energy and X-ray energy are statistically independent, such that the joint probability distribution factorizes into the product of the individual energy spectra.
Although the remaining background contains both accidental and true physical coincidences, the background density in the vicinity of the signal region is observed to be approximately separable in $\gamma$-ray and X-ray energy.

The number of coincident signal events is found in the $\gamma$-ray energy interval [561\,keV, 565\,keV] and the X-ray energy interval [9.46\,keV, 10.24\,keV] marked as $A$ in figure~\ref{fig:CoincProj}. 
The background $A_\mathrm{BG}$ contained in the signal region is quantified with a two-dimensional sideband approach. 
If the two-dimensional probability distribution of background energies factorizes into the individual spectra as $P(E_\gamma, E_\mathrm{X-ray}) = P(E_\gamma) P(E_\mathrm{X-ray})$, then 
the proportion $A_\mathrm{BG}/B = D/C$ holds. 
$A_\mathrm{BG}$ is the number of true and random coincident background events contained in the signal region and $B$, $C$ and $D$ are coincident counts with different $\gamma$-ray and X-ray energy cut conditions. 
$B$ stands for events fulfilling the X-ray signal cut and the $\gamma$-ray background cut. $D$ stands for events fulfilling the X-ray background cut and the $\gamma$-ray signal cut. $C$ stands for both fulfilling the background cut condition, as labeled on figure~\ref{fig:CoincProj}. 
The $\gamma$-ray background energy interval is [565\,keV, 1800\,keV]. The X-ray background energy region consists of the intervals [2\,keV, 9.46\,keV] and [12\,keV, 20\,keV], excluding the As $K_\alpha$ and $K_\beta$ X-rays at 10.5\,keV and 11.7\,keV. 
The background regions are chosen to maximize the available statistics while excluding regions containing signal events or strong correlated structures from characteristic X-rays such that the two-dimensional sideband method stays applicable. Since the background estimate relies on the separability assumption, the selected intervals are validated using the consistency checks shown in figure~\ref{fig:Stat_indep_cross_check} and described below.
The coincident EC$^\ast$ signal net counts are calculated as 
\begin{equation}
N_\mathrm{coinc} = A - A_\mathrm{BG} = A - \frac{BD}{C}.
\end{equation}
Counting statistics in the number of coincident events introduces a dominant statistical uncertainty of \SI{4}{\percent}.

\subsection{Branching ratio calculation}
The $\gamma$-ray line at 559.1\,keV with the emission probability $(40.670\pm0.291)\%$ \cite{Marnada2000} is used as $^{76}$As activity reference. It is evaluated in the HPGe spectrum without any coincidence cuts applied. A cut and count approach is applied with a summation window [557.1\,keV, 561.1\,keV]. It has the same energy width as the $\gamma$-ray energy cut used for the coincident counts. The background sidebands [551.1\,keV, 553.1\,keV] and [565.1\,keV, 567.1\,keV] are symmetrical on both sides while the upper sideband is not affected by the 563.1\,keV satellite peak. 

The EC$^\ast$ branching ratio is calculated as 
\begin{equation}
\nu_{\mathrm{EC}^\ast} = 
\frac{
    N_\mathrm{coinc} \,
    t_\mathrm{real} \,
    k_\mathrm{RCS}
}{
    t_\mathrm{live}^\mathrm{SDD}
}
\;
\frac{1}
{
    p_K \,
    \omega_K \,
    p_{\alpha} \,
    \varepsilon_\mathrm{coinc}
} 
\;
\frac{
    \nu_\mathrm{ref} \,
    \varepsilon_\mathrm{ref}
}
{
    N_\mathrm{ref}
}
\label{eq:nu_EC}
\end{equation}
with the parameters summarized in table~\ref{tab:nu_EC_par}.
\begin{table}[b]
    \caption{Parameters of equation~\ref{eq:nu_EC} with the values for the whole dataset. 
    \label{tab:nu_EC_par}
    }
    \begin{ruledtabular}
        \begin{tabular}{l l r}
            \multicolumn{1}{l}{Symbol} & Meaning & \multicolumn{1}{r}{Value} \\
            \hline
            $N_\mathrm{coinc}$ & Coincident signal counts & $740\!\pm\!34$ \\
            $t_\mathrm{real}$  & Real time & 741.8\,h \\ 
            $t_\mathrm{live}^\mathrm{SDD}$ & X-ray detector live time & 144.8\,h \\ 
            $k_\mathrm{RCS}$ & X-ray summing correction & 1.020$^\ast$ \\ 
            $p_K$ & $K$-capture ratio & $0.877\!\pm\!0.009$ \cite{Turkat2023} \\
            $\omega_K$ & Ge $K$ fluorescence yield & $0.546\!\pm\!0.004$ \cite{Schoenfeld1996} \\
            $p_\alpha$ & $K_\alpha$ emission probability & $0.869\!\pm\!0.009$ \cite{Schoenfeld2000} \\
            $\varepsilon_\mathrm{coinc}$ & Coincidence efficiency & $(0.052\!\pm\!0.007)\%$ \\
            $N_\mathrm{ref}$ & 559.1\,keV counts & $(6.93910\!\pm\!0.00026)\!\times\!10^8$\\
            $\nu_\mathrm{ref}$ & \SI{559.1}{\keV} em. prob. & $(40.670\!\pm\!0.291)\%$ \cite{Marnada2000} \\
            $\varepsilon_\mathrm{ref}$ & HPGe efficiency & $(5.0\!\pm\!0.5)\%$
        \end{tabular}
    \end{ruledtabular}
    \raggedleft
    $^\ast$effective value
\end{table}
$N_\mathrm{coinc}$ is the number of coincident signal counts in the peak at $E_\gamma = 562.9\,$keV and $E_{\text{X-ray}} = 9.9\,$keV. Using this experimental input, the first two fractions evaluate the number of EC$^\ast$ decays having occurred in the sample during the measurement. 
$t_\mathrm{real}$ is the real time and
$t_\mathrm{live}^\mathrm{SDD}$ the live time of the X-ray detector. 
$k_\mathrm{RCS}$ is a correction factor for X-ray summing-out by random coincidence summing. Random summing in the HPGe affects both $N_\mathrm{coinc}$ and $\nu_\mathrm{ref}$ in the same way and thus cancels out. 
The X-ray detector dead time and random summing corrections are applied time-resolved because they are depending on the count rate which is changing significantly during the measurements due to the decay of $^{76}$As. The first term in equation~\ref{eq:nu_EC} is evaluated for 1-hour time intervals of the measurement and then summed. 
The dead time ranges from 79\% for the largest count rates to 73\% for the lowest count rates. 
The second term corrects for the probability of having full-energy depositions in both detectors at the signal energies given an EC$^\ast$ decay occurs in the sample. 
$p_K = 0.877\pm0.009$ \cite{Turkat2023} stands for the probability of $K$-capture given an EC$^\ast$ happens. 
$\omega_K = 0.546\pm0.004$ is the $K$ vacancy fluorescence yield of germanium \cite{Schoenfeld1996}, 
$p_\alpha = 0.869\pm0.018$ the $K_\alpha$ emission probability given a $K$ X-ray emission \cite{Schoenfeld2000}, and
$\varepsilon_\mathrm{coinc} = (0.052\pm0.007)\%$ the efficiency of the 2-detector coincidence setup for $E_\gamma = 562.9\,$keV and $E_{\mathrm{X-ray}} = 9.9\,$keV. 
The last term in equation~\ref{eq:nu_EC} is the inverse of the number of $^{76}$As decays that took place in the sample during the measurement. 
$N_\mathrm{ref}$ is the background-subtracted number of counts in the HPGe reference peak at 559.1\,keV, 
$\nu_\mathrm{ref} = (40.670\pm0.291)\%$ the emission probability of the \SI{559.1}{\keV} reference $\gamma$-ray, and 
$\varepsilon_\mathrm{ref} = (5.0\pm0.5)\%$ the HPGe full-energy efficiency for \SI{559.1}{\keV} $\gamma$-ray detection. For the latter, the typical systematic uncertainty in absolute efficiency simulations of 10\% is assumed. 

The X-ray detector reset preamplifier output voltage range exceeded the DAQ input range by a factor of three. 
The rate and energy dependent live time was estimated taking into account the preamplifier sawtooth shape and the simulated spectrum. Test experiments with known live times are compared to this estimate. Their deviation of \SI{4.2}{\%} is assumed to describe the uncertainty of the estimated live time. The dead time estimation affects the $^{76}$As EC$^\ast$ measurements with varying count rates and the X-ray detector calibration with constant count rates. A rate-independent bias in the dead time correction factor estimate would be canceled out. 
By variation of the live time estimate in both calibration and EC$^\ast$ measurements, it is found that the uncertainty of the branching ratio introduced by the dead time estimation is 2\%. 

The detection efficiencies are taken from a Geant4 \cite{Agostinelli2003} particle transport simulation of the setup. 
The implementation of the detectors was optimized to match calibration measurements with point-like radiation standards. For the X-ray detector, sources of $^{57}$Co, $^{133}$Ba and $^{65}$Zn were used at a distance of \SI{50}{\mm} from the detector. 
For the HPGe, calibration measurements with standards of $^{137}$Cs, $^{88}$Y, $^{60}$Co and $^{57}$Co at distances 0.25\,cm, 3\,cm and 7\,cm were used. 
The efficiencies ratio found this way is $\varepsilon_\mathrm{coinc} / \varepsilon_\mathrm{ref} = 0.0102\pm0.0013$. 
The effects of photon self absorption in the sample and true coincidence summing are included in the simulated efficiencies. 

The largest systematic uncertainty is introduced by the detection efficiency calibration of the X-ray detector. The maximum deviation between a calibration line's simulated and experimental efficiency of 12\,\% is assumed to describe the X-ray Monte-Carlo systematic uncertainty. 
In the detection efficiencies term in equation~\ref{eq:nu_EC}, the coincidence detection efficiency $\varepsilon_\mathrm{coinc}$ can be expressed as the product of both detectors' full energy detection efficiencies for the respective signal energies \SI{562.9}{\keV} and \SI{9.9}{\keV}: 
\begin{equation}
    \frac{\varepsilon_\mathrm{ref}}{\varepsilon_\mathrm{coinc}}
    = \frac{1}{\varepsilon_\mathrm{SDD}(9.9\,\mathrm{keV})}\cdot
    \frac{\varepsilon_\mathrm{HPGe}(559.1\,\mathrm{keV})}{\varepsilon_\mathrm{HPGe}(562.9\,\mathrm{keV})}
    \label{eq:cancel_eff}
\end{equation}
The effect of the HPGe geometric acceptance cancels out. 
The remaining ratio of full energy detection probabilities for the two $\gamma$-ray energies close to each other has a 
minor systematic uncertainty compared to the X-ray detection efficiency Monte-Carlo systematic uncertainty. 
Remaining systematic uncertainties in the ratio of HPGe efficiencies are the efficiency energy dependence and the Geant4 description of true coincidence summing. They are estimated to be 0.04\% and 0.4\%, respectively, and are considered positively correlated. 
Systematic uncertainty contributions are summarized in table~\ref{tab:syst_unc}. 
\begin{table}[]
    \caption{Systematic uncertainty included in parameters in equation~\ref{eq:nu_EC} and their impact on the the EC$^\ast$ branching ratio result. 
    \label{tab:syst_unc}
    }
    \begin{ruledtabular}
        \begin{tabular}{lrlr}
            Parameter & Syst. & Assumption & Eff. on BR \\
            \hline
            $\varepsilon_\mathrm{SDD}$      & 12\%   & Max. dev. sim./meas.  & 12\%   \\ 
            $\varepsilon_\mathrm{HPGe}$     & 10\%   & Cancel geom.          & 0.44\% \\
            $t_\mathrm{live,SDD}$           & 4.2\%  & Dev. from test exp.   & 2\%    \\
            $\nu_\mathrm{ref}$              & 0.74\% & \cite{Marnada2000}    & 0.74\% \\
            $\omega_K$                      & 0.73\% & \cite{Schoenfeld1996} & 0.73\% \\
            $p_\alpha$                      & 2.1\%  & \cite{Schoenfeld2000} & 2.1\% \\
            $p_K$                           & 1.0\%  & \cite{Turkat2023}     & 1.0\% \\
        \end{tabular}
    \end{ruledtabular}
\end{table}
Considering all given contributions uncorrelated, the relative systematic uncertainty of the measured EC$^\ast$ branching ratio is 13.0\%.  

\begin{table*}
\caption{
    Run-wise evaluation of the $^{76}$As EC$^\ast$ branching ratio. 
    Eleven activations were done each followed by a coincidence measurement. Three measurements were interrupted leading to two measurement runs after one activation. 
    $R$ is the count rate in the HPGe. 
    Coincident signal counts $A$ and coincident background counts $B$, $C$ and $D$ are evaluated as shown in figure~\ref{fig:CoincProj} and used for the coincident net counts calculation. $K_{t\mathrm{-dep.}} = k_\mathrm{RCS} \cdot t_\mathrm{real} / 
    t_\mathrm{live}^\mathrm{SDD}$
    is the correction factor for rate-dependent effects including the dead time correction $t_\mathrm{real} / t_\mathrm{live,SDD}$ and the random coincidence summing correction for the X-ray detector both evaluated in a 1-hour time binning. Reference counts are evaluated from the peak in the HPGe spectrum at $E_\gamma = 559.1\,$keV. The branching ratio (BR) is calculated for each measurement run. The statistical uncertainty is based on the uncertainties of the coincidence net counts and the reference net counts. }
\begin{ruledtabular}
\begin{tabular}{r r r r r r r r r r r r r}
Act. & Run & $t_\mathrm{real}$/h & \multicolumn{2}{c}{$R$ ($\times\!10^3\,\mathrm{s}^{-1}$)} & \multicolumn{5}{c}{Coincident counts} & $K_{t\mathrm{-dep.}}$ & Ref. cts. & BR in \%\\
  &  &  & max. & min. & $A$ & $B$ & $C$ & $D$ & $N_\mathrm{coinc}$ &  & ($\!\times\!10^6$) & stat. unc. \\
\hline 
I	& 1	 & 63	& 3.0   & 0.6	& 87	& 514	& 5059	& 240	& 63$\pm$9	& 5.08	& 48.100$\pm$0.007	& 0.063$\pm$0.009 \\
II	& 2	 & 136	& 3.8	& 0.1	& 123	& 688	& 7719	& 391	& 88$\pm$11	& 5.10	& 74.040$\pm$0.009	& 0.058$\pm$0.007 \\
III	& 3	 & 63	& 3.3	& 0.6	& 98	& 612	& 6211	& 346	& 64$\pm$10	& 5.46	& 53.161$\pm$0.007	& 0.063$\pm$0.010 \\
IV	& 4	 & 38	& 4.5	& 1.7	& 102	& 709	& 7059	& 460	& 56$\pm$10	& 6.34	& 56.681$\pm$0.008	& 0.060$\pm$0.011 \\
V	& 5	 & 15	& 5.5	& 3.7	& 60	& 491	& 4861	& 327	& 27$\pm$8	& 8.36	& 35.694$\pm$0.006	& 0.060$\pm$0.017 \\
	& 6	 & 94	& 3.5	& 0.3	& 109	& 649	& 6579	& 381	& 71$\pm$10	& 5.10	& 63.014$\pm$0.008	& 0.055$\pm$0.008 \\
VI	& 7	 & 18	& 4.5	& 2.8	& 57	& 388	& 3930	& 208	& 36$\pm$8	& 6.21	& 32.897$\pm$0.006	& 0.066$\pm$0.014 \\
	& 8	 & 18	& 2.6	& 1.6	& 32	& 185	& 1955	& 117	& 21$\pm$6	& 5.31	& 19.696$\pm$0.005	& 0.054$\pm$0.015 \\
VII	& 9	 & 87	& 5.3	& 0.6	& 150	& 1095	& 10835	& 584	& 91$\pm$12	& 5.89	& 94.028$\pm$0.010	& 0.054$\pm$0.007 \\
VIII& 10 & 39	& 4.1	& 1.5	& 92	& 596	& 5890	& 316	& 60$\pm$10	& 5.81	& 52.238$\pm$0.007	& 0.064$\pm$0.010 \\
IX	& 11 & 16	& 3.1	& 2.0	& 26	& 203	& 2228	& 106	& 16$\pm$5	& 5.59	& 20.548$\pm$0.005	& 0.042$\pm$0.013 \\
	& 12 & 19	& 1.8	& 1.1	& 23	& 121	& 1308	& 90	& 15$\pm$5	& 4.82	& 14.020$\pm$0.004	& 0.048$\pm$0.016 \\
X	& 13 & 65	& 4.4	& 0.8	& 114	& 808	& 7835	& 444	& 68$\pm$11	& 5.74	& 71.001$\pm$0.009	& 0.053$\pm$0.008 \\
XI	& 14 & 70	& 3.5	& 0.6	& 100	& 617	& 6012	& 360	& 63$\pm$10	& 5.08	& 58.794$\pm$0.008	& 0.052$\pm$0.008 \\
\hline
\multicolumn{2}{r}{1-14} & 742   & 5.5   & 0.1   & 1173  & 7676  & 77481 & 4370  & 740$\pm$34 & 5.62  & 693.910$\pm$0.027 & 0.0572$\pm$0.0029 \\

\end{tabular}
\end{ruledtabular}
\label{tab:runs}
\end{table*}
The extracted parameters for all measurement runs are summarized in table~\ref{tab:runs}. 
The measured $^{76}$As EC$^\ast$ branching ratio using the total data set is $\nu_{\mathrm{EC}^\ast} = 0.0572\%$ as given in the last row. It is calculated using the summed counts rather than averaging the branching ratio measured in each run. 
A mean weighted according to the inverse uncertainties would prefer statistically fluctuated smaller coincident counts. 

\subsection{Cross checks}

\begin{figure}
    \centering
    \includegraphics[width=\linewidth]{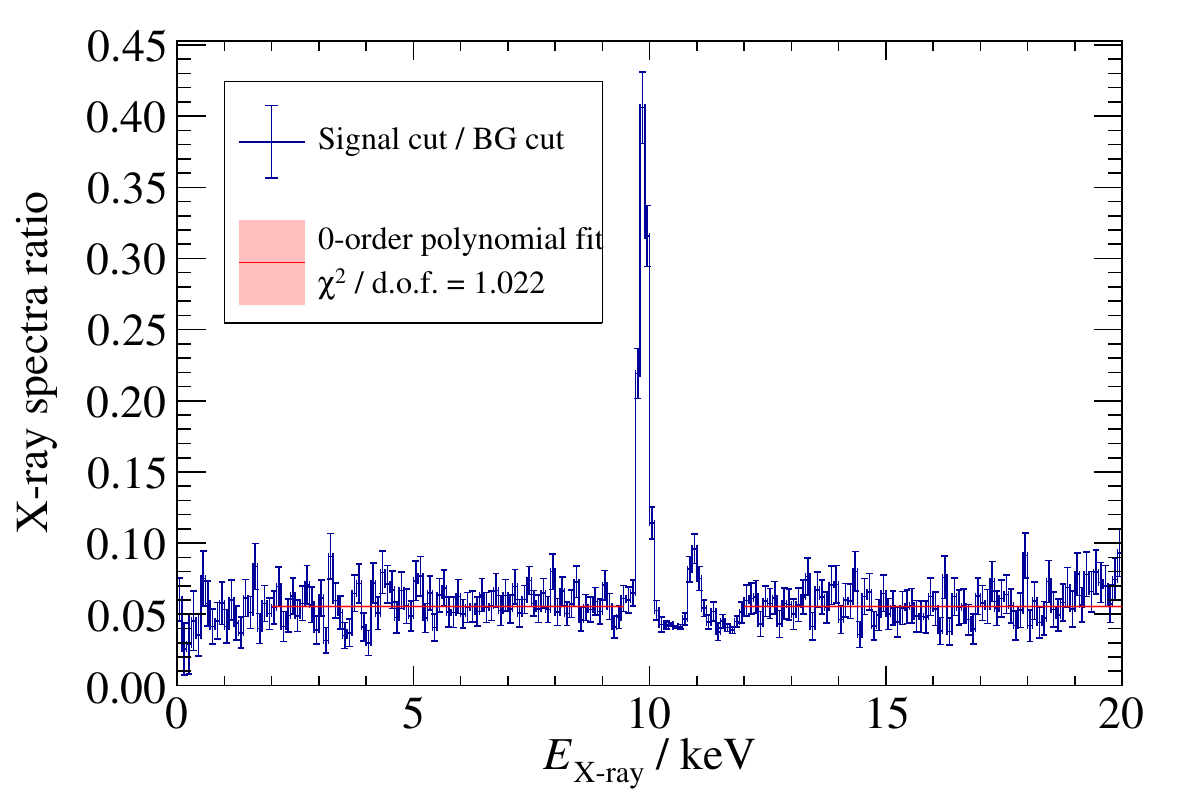}
    \includegraphics[width=\linewidth]{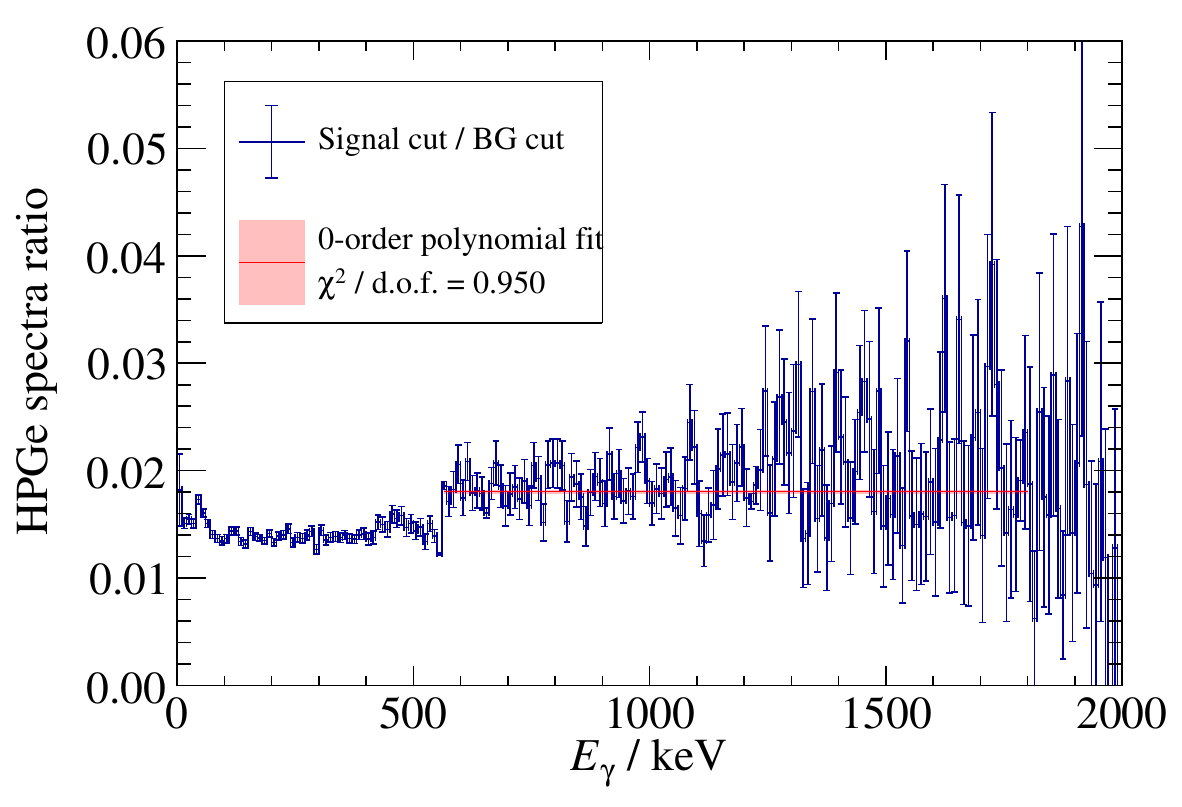}
    \caption{Cross-check of statistical independence of coincident backgrounds for the whole dataset. Top: The X-ray spectrum with the EC$^\ast$ signal $\gamma$-ray coincidence cut condition $561\,\mathrm{keV}\!<\!E_\gamma\!<\!565\,\mathrm{keV}$ is divided by the one with the background (BG) $\gamma$-ray coincidence cut condition $565\,\mathrm{keV} < E_\gamma\!<\!1800\,\mathrm{keV}$. 
    Bottom: The $\gamma$ spectrum with the EC signal X-ray coincidence cut condition $9.46\,\mathrm{keV}\!<\!E_\mathrm{X-ray}\!<\!10.24\,\mathrm{keV}$ is divided by the one with the background X-ray coincidence cut condition $2\,\mathrm{keV}\!<\!E_\mathrm{X-ray}\!<\!9.46\,\mathrm{keV}$ or $12\,\mathrm{keV}\!<\!E_\gamma\!<\!20\,\mathrm{keV}$. }
    \label{fig:Stat_indep_cross_check}
\end{figure}

The applicability of the two-dimensional sideband method
is experimentally tested in figure~\ref{fig:Stat_indep_cross_check}. If $\gamma$-ray and X-ray background energies are independent, the ratio between signal and background projections should be constant outside the signal peaks. 
This behavior is observed. 
The number of coincident signal counts is cross-checked with 1-dimensional peak fits in both projections and a two-dimensional fit. 
A partitioned approach gives results consistent with an evaluation of the summed spectra of all measurements for all mentioned methods.

\begin{figure}[h]
    \centering
    \includegraphics[width=\linewidth]{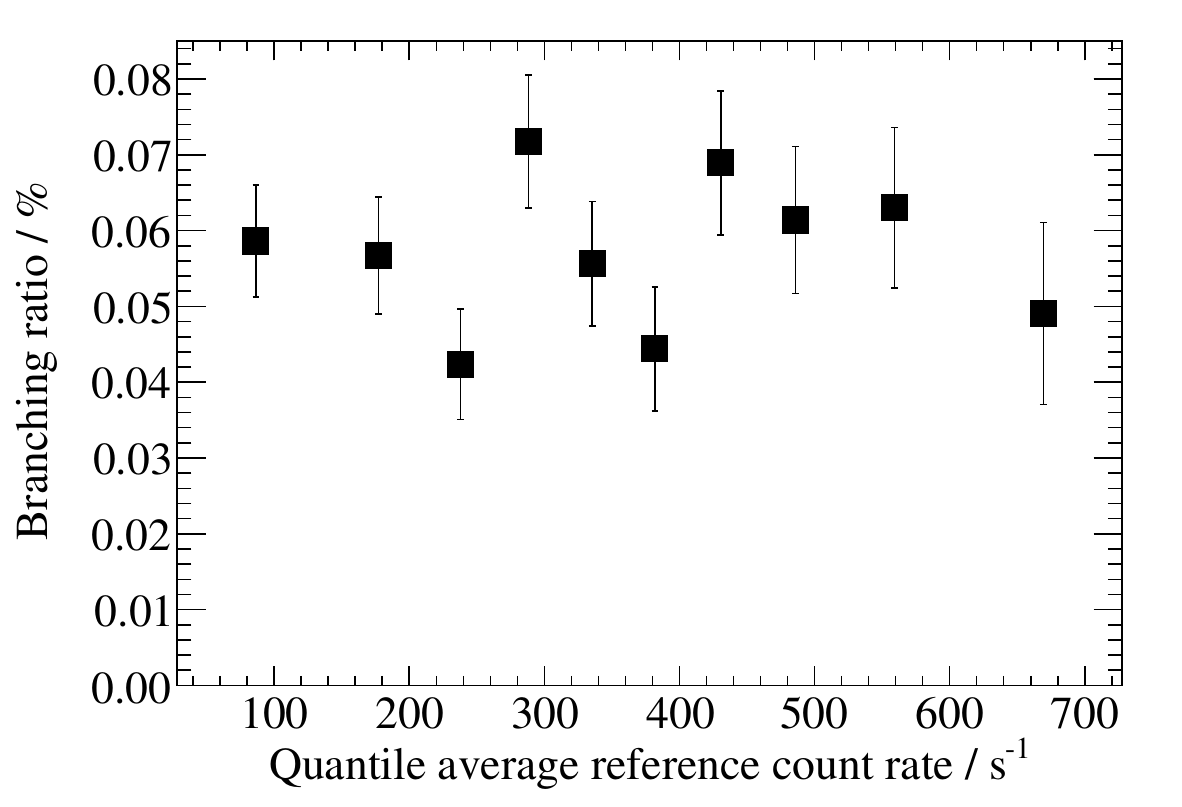}
    \caption{Cross-check on the rate dependence of the data analysis. Data are sorted into ten quantiles by the \SI{559.1}{\keV} reference count rate in the HPGe such that each quantile contains the same number of reference counts. With all rate dependent corrections applied, this is consistent with a constant. 
    }
    \label{fig:RateQuantileBR}
\end{figure}

The count rates in both detectors spread over a wide range in the experiment. 
The extracted result strongly depends on the correct description of time-dependent corrections. 
This includes the dead time, random summing and even a potential peak shape dependence on the count rate. 
A cross check for time-dependent systematics is done by investigation of the influence of the count rate on the calculated branching ratio. The measurement is separated into time intervals of one hour each. The time interval sub-measurements are sorted by the count rate of the \SI{559.1}{\keV} reference line in the HPGe. Any count rate dependence would appear as a trend in the measured branching ratio over the count rate. In order to have convincing statistics, data are summarized by rate into eight quantiles. The measured branching ratio is shown in figure~\ref{fig:RateQuantileBR} over the average count rate of the respective quantile. 
The individual branching ratios calculated from each rate quantile data set are consistent with each other considering their statistical uncertainties. Thus, independence from the count rate is assumed. 

The energy calibration and $\gamma$ peak shapes are found changing between groups of runs but constant within runs. 
Dependence on these parameters is cross-checked by comparison to a partitioned approach. Three partitions are defined such that in each the energy calibration and peak shape parameters are constant. Results found when evaluating partitions are consistent with the evaluation of spectra summed for the whole dataset. 

\section{Conclusion}
The electron capture of $^{76}$As into the first excited state of $^{76}$Ge was measured for the second time. 
The measured branching ratio
\begin{equation}
    \nu_{\mathrm{EC}^\ast} = (0.0572 \pm 0.0029 (\mathrm{stat.}) \pm 0.0074(\mathrm{syst.}))\%
\end{equation}
is in the order of the theory predictions of $\approx$0.03\% and 70\% larger than the previous measurement of $\approx$0.027\% \cite{Domula2014}. 
For the first time the full uncertainty budget was quantified. 
The measurement is presently limited by the uncertainty of the X-ray detection efficiency. 
The systematic uncertainty is in the same order as in the previous measurement while the relative statistical uncertainty is by a factor of 6 lower. 

The EC$^\ast$ branching ratio measured in this work translates into $\mathrm{log}\,ft = 6.99$ \cite{Turkat2023}. 

\subsection{EC into the \textsuperscript{76}Ge ground state}
The present measurement provides the first quantified uncertainty budget for the EC$^\ast$ branching ratio of $^{76}$As and establishes the experimental input required for future investigations of the remaining unmeasured EC branch. Further constraints on nuclear structure calculations relevant for double-beta decay can be obtained by measuring the EC$^0$ transition, the direct decay into the $^{76}$Ge ground state. A future measurement of this branch requires separation from the EC$^\ast$ contribution measured in this work. Since both transitions produce identical characteristic Ge X-rays, discrimination requires either suppression of the accompanying 562.9\,keV $\gamma$ ray from the EC$^\ast$ decay or an independent measurement of the total EC probability.

A $\gamma$-ray anti-coincidence detector should have a large total detection efficiency $\varepsilon_\mathrm{veto}$ for the 562.9\,keV $\gamma$-rays emitted in the EC$^\ast$ because the EC$^0$ sensitivity scales with 
$(1-\varepsilon_\mathrm{veto})$. The time resolution should be not worse than that of the X-ray detector while energy resolution is less relevant. These demands are fulfilled by a scintillation detector surrounding the setup of $^{76}$As source and X-ray detector. The scintillation volume has to be large enough to absorb a sufficient portion of the 562.9\,keV $\gamma$-rays emitted by the sample. 
This is the approach of the KDK collaboration for the measurement of the $^{40}$K EC$^0$ decay \cite{Hariasz2023}. A veto detector with the dimensions of the Modular Total Absorption Spectrometer \cite{Karny2016} that was used to veto the 1460.8\,keV $\gamma$-rays emitted by $^{40}$K provides an excellent sensitivity to the $^{76}$As EC$^0$. 
A sufficient sensitivity considering the expected ratio of EC$^0$/EC$^\ast$ in $^{76}$As can be achieved with less efforts. The $\gamma$-ray veto detector can be smaller thanks to the shorter absorption length. 
The scintillation detector could be, e.g., a benzene liquid scintillator volume 120\,cm in all dimensions, with the source in its center, and the X-ray detector reaching into the liquid close to the source. The liquid scintillator is optically coupled to  multiple photomultiplier tubes mounted on the outside. 
The 60\,cm of benzene surrounding the source 
transmit less than 1\% of the 562.9\,keV $\gamma$-rays. 
Geant4 simulations of such a setup assuming a low energy threshold of 100\,keV for the scintillation veto detector suggest a 562.9\,keV $\gamma$-ray veto efficiency of 98.4\% and a reduction of the continuous background in the X-ray detector to 26\%. 

In the present work it was shown that in a silicon drift detector measuring the Ge X-rays there is a dominant continuous background caused by $\beta^-$ particles. In order to measure the total EC with an SDD, this background has to be suppressed with an active $\beta$ veto detector, e.g. an organic scintillator between sample and X-ray detector measuring transmitted $\beta^-$ particles while transmitting sufficient Ge X-rays. 

Atom counting can be done on an As sample, activated via $^{75}$As(n,$\gamma$) with the thermal neutrons provided by a reactor, after the decay of the produced $^{76}$As for several half-lives. The ratio of atoms $^{76}$Ge/$^{76}$Se can be measured via mass spectrometry. An extremely pure As sample is needed and a sufficiently large neutron fluence has to be applied in order to produce significantly more $^{76}$Ge than the Ge contamination contained in the As sample. A purification of AsCl$_3$ to the grade 7N has been achieved \cite{Zou2026}. 

Summarizing, different experimental setups are possible for the measurement of the electron capture of $^{76}$As into the ground state of $^{76}$Ge. 
Building upon this work, an anti-coincidence setup consisting of a SDD and a $4\pi$ scintillation detector 
offers the best prospects for this unprecedented measurement. 

\section{Acknowledgments}

We thank J.~Suhonen for fruitful discussions and for providing theoretical estimates on the branching ratio. 
We thank C.~Lange and D.~Gehre 
for help operating the TU~Dresden AKR-2 reactor, and C.~Seibt for experimental support. 
This work was supported by the Bundesministerium f\"{u}r Bildung und Forschung (BMBF-Projekte 05A20OD1 und 05A23OD1).

%
\def\AAA{Astro. and Astrophys.}
\def\AAS{Astro. and Astrophys. Suppl.}
\def\AHEP{Adv. High Energy Phys.}
\def\AIPCP{AIP Conf. Proc.}
\def\ANDT{Atom. Nucl. Data Tab.}
\def\ANDT{Atomic and Nuclear Data Tables}
\def\ADNDT{At. Data Nucl. Data Tables}
\def\ANE{Ann. Nucl. Energy}
\def\APB{{Acta Pol.} B}
\def\APJ{ApJ}
\def\arxiv{arXiv}
\def\APJ{The Astrophys. J.}
\def\APJL{The Astrophys. J. Lett.}
\def\APP{Astropart. Phys.} 
\def\ARI{Appl. Rad. Isot.}
\def\ARNPS{Ann. Rev. Nucl. Part. Sci.}
\def\CJP{Can. J. Phys.}
\def\CPC{Chin. Phys. C}
\def\EPJA{Europ. Phys. J. A}
\def\EPJC{Europ. Phys. J. C}
\def\EPJP{Europ. Phys. J. Plus}
\def\EPJWC{EPJ Web conf.}
\def\EPL{Europhys. Lett.}
\def\ITNS{IEEE Trans. Nucl. Sci.}
\def\JCAP{JCAP}
\def\JINST{JINST}
\def\JPCS{J. Phys. Conf. Ser.}
\def\JPG{J. Phys. G}
\def\MAT{Materials}
\def\MNRAS{Month Not. Royal Ast. Soc.}
\def\MUP{Moscow Univ. Phys.}
\def\NAF{Z. Naturforschung A} 
\def\NAT{Nature}
\def\NCI{Nuovo Cimento C}
\def\NDS{Nucl. Data Sheets}
\def\NIM{Nucl. Instrum. Methods}
\def\NIMA{{Nucl. Instrum. Methods} A}
\def\NP{Nucl. Phys.}
\def\NPA{{Nucl. Phys.} A} 
\def\NPB{{Nucl. Phys.} B} 
\def\NPBP{{Nucl. Phys.} B (Proc. Suppl.)} 
\def\PAC{Pure Appl. Chem.}
\def\PAN{{Physics of Atomic Nuclei}} 
\def\PHM{Philos. Mag.} 
\def\PLB{{Phys. Lett.} B}
\def\PNpp{Prog. Nucl. Part. Phys.}  
\def\PR{Phys. Rev.} 
\def\PRC{{Phys. Rev.} C} 
\def\PRD{{Phys. Rev.} D} 
\def\PRL{Phys. Rev. Lett.} 
\def\PRP{{Phys. Rep.}}
\def\PTEP{Prog. Th. Exp. Phys.}
\def\PTP{{Prog. Theo. Phys.}}
\def\RMP{{Rev. Mod. Phys.}}
\def\Rpp{Rep. Prog. Phys.}
\def\SCI{Science} 
\def\ZP{Z. Phys.} 
\def\ZPA{{Z. Phys.} A} 
\def\ZPC{{Z. Phys.} C} 

\end{document}